\documentclass[journal,twoside,web]{ieeecolor}
\usepackage{tmi}
\usepackage{cite}
\usepackage{textcomp}
\usepackage{cite}
\usepackage{float}

\usepackage[pdftex]{graphicx}

 \DeclareGraphicsExtensions{.eps}

\usepackage{subcaption}

\usepackage{amsmath,amssymb,amsfonts}
\usepackage{bm}
\usepackage{diagbox}

\usepackage{balance}

\usepackage{algorithmic}

\usepackage{lipsum}
\usepackage{footnote}
\usepackage{hyperref}
\usepackage{stfloats}
\usepackage{authblk}

\usepackage{booktabs}       

\usepackage{url}
\usepackage{threeparttable}
\usepackage{multirow}
\usepackage{makecell}
\usepackage{tabularx}
\usepackage{color}

\def\BibTeX{{\rm B\kern-.05em{\sc i\kern-.025em b}\kern-.08em
    T\kern-.1667em\lower.7ex\hbox{E}\kern-.125emX}}
\begin{document}
\title{An Interpretable Cross-Attentive Multi-modal MRI Fusion Framework for Schizophrenia Diagnosis}
\author{Ziyu Zhou, Anton Orlichenko, Gang Qu, Zening Fu, Vince D Calhoun, \IEEEmembership{Fellow, IEEE}, Zhengming Ding and Yu-Ping Wang,
 \IEEEmembership{Senior Member, IEEE}
\thanks{This work was partially supported by NIH Grants R01 GM109068, R01 MH104680, R01 MH107354, P20 GM103472, R01 REB020407, R01 EB006841, NSF Grant \#1539067. Corresponding author: Yu-Ping Wang, wyp@tulane.edu}
\thanks{Ziyu Zhou and Zhengming Ding are with the Department of Computer Science, Tulane University, New Orleans, LA, 70118, USA. }
\thanks{Anton Orlichenko, Gang Qu, Yu-Ping Wang are with the Department of Biomedical Engineering, Tulane University, New Orleans, LA, 70118, USA. }
\thanks{Zening Fu and Vince D Calhoun are with Tri-Institutional Center for Translational Research in Neuroimaging and Data Science (TReNDS), (Georgia State University, Georgia Institute of Technology, Emory University), Atlanta, GA, 30303, USA.}}

\maketitle

\begin{abstract}
Both functional and structural magnetic resonance imaging (fMRI and sMRI) are widely used for the diagnosis of mental disorder. However, combining complementary information from these two modalities is challenging due to their heterogeneity. Many existing methods fall short of capturing the interaction between these modalities, frequently defaulting to a simple combination of latent features. In this paper, we propose a novel Cross-Attentive Multi-modal Fusion framework (CAMF), which aims to capture both intra-modal and inter-modal relationships between fMRI and sMRI, enhancing multi-modal data representation. Specifically, our CAMF framework employs self-attention modules to identify interactions within each modality while cross-attention modules identify interactions between modalities. Subsequently, our approach optimizes the integration of latent features from both modalities. This approach significantly improves classification accuracy, as demonstrated by our evaluations on two extensive multi-modal brain imaging datasets, where CAMF consistently outperforms existing methods. Furthermore, the gradient-guided Score-CAM is applied to interpret critical functional networks and brain regions involved in schizophrenia. The bio-markers identified by CAMF align with established research, potentially offering new insights into the diagnosis and pathological endophenotypes of schizophrenia.

\end{abstract}


\section{Introduction}\label{sec:introduction}
Medical imaging powered by machine learning has been widely explored in the diagnosis of mental disorders such as schizophrenia \cite{oh2020identifying}, autism \cite{katuwal2015predictive}, and Alzheimer's disease (AD) \cite{tomassini2021end}. Notably, machine learning models aid in the identification of biomarkers, including critical brain regions associated with specific diseases. Both functional magnetic resonance imaging (fMRI) and structural magnetic resonance imaging (sMRI) have emerged as popular imaging modalities for mental disorder diagnosis. fMRI offers insights on the functional organization of the brain by quantifying the changes in blood-oxygen-level-dependent (BOLD) signal of the human brain \cite{belliveau1991functional}. The functional connectivity (FC) derived from fMRI depicts the correlation between brain regions of interest (ROIs), and has proven powerful in age prediction \cite{li2018brain} and disease diagnosis \cite{zhang2017hybrid}. Many research works \cite{wang2021functional, finn2015functional} demonstrate that FC can be viewed as unique ``brain fingerprints'', differentiating individuals with varying cognitive and behavioral abilities. sMRI, on the other hand, captures structural information about the human brain including morphology (e.g., surface area, cortical thickness, etc.). Structural changes in certain brain regions can be linked to specific phenotypical traits. Despite their individual strengths, combining the complementary information from sMRI and fMRI to improve disease diagnosis remains challenging, largely due to the heterogeneity of these two modalities.


Along this line, various methods have been proposed to integrate heterogeneous data from multiple modalities. Multi-set canonical correlation analysis (MCCA) was proposed \cite{sui2013three} for the fusion of multi-modality brain image data such as fMRI, sMRI, and diffusion MRI (dMRI). Zu et al. (2016) \cite{zu2016label} used a label-aligned regularization term in a multi-kernel learning framework to utilize the relationships between multiple modalities for the selection of important feature subsets with the result of improving the model performance on AD classification. However, many existing data fusion methods don't fully utilize the potential synergy between these modalities, frequently relying on simple concatenation of latent features. This limitation restricts the potential to fully showcase the enhanced utility derived from multi-modal data integration. On the other hand, advancements in multi-modal deep learning models offer promising solutions to the challenge of modality fusion. Yang et al. (2023) \cite{yang2023mapping} utilized graph-based networks to fuse multi-modal brain images, achieving promising performance. Hu et al. (2019) \cite{hu2019deep} proposed a deep collaborative learning (DCL) method that incorporated the linear and non-linear correlations of two datasets, addressing the limitations of canonical correlation analysis (CCA), a linear model. Following this, Hu et al. \cite{hu2021interpretable} proposed a collaborative layer to integrate latent features from single nucleotide polymorphism (SNPs) and FCs and improved the ability of their model to predict general fluid intelligence.

Recently, the transformer architecture witnessed great success with the capability of the attention mechanism to extract interaction between elements in a sequence \cite{vaswani2017attention}. The attention mechanism shows great potential for adaptation to our task of fusing the latent features from fMRI and sMRI, as it can discover interrelations in heterogeneous data across various features \cite{qu2023interpretable}. Therefore, in this work, we propose incorporating the attention mechanism into a two-level data fusion framework. At the first level, we use self-attention (SA) modules to extract interactions within each modality while cross-attention (CA) \cite{zhu2022multimodal} modules to explore interactions across fMRI and sMRI. At the second level, we integrate the latent features from these four attention pathways, dynamically updating weights throughout the model's training process to achieve optimal integration.


Interpretable AI is crucial for correct comprehension of machine learning model decisions, especially for computer-assisted diagnosis \cite{salahuddin2022transparency, huff2021interpretation}. Methodologically, the interpretability of machine learning models can be categorized as backpropagation-based and perturbation-based types. The former, gradient-based, assesses input feature influence through neural network backpropagation. Conversely, perturbation-based methods alter inputs to observe output variations for explanatory purposes. Among representative works, the gradient-weighted class activation mapping (Grad-CAM) method \cite{selvaraju2017grad} was proposed as an alternative to calculate the weights for feature maps from the gradients of the output nodes via backpropagation. This design allowed for a more flexible architecture without necessitating modifications to the underlying model structure. Score-CAM \cite{wang2020score} further advanced this method, which generated the feature map weights by measuring the performance change after masking the input with each feature map. The heatmap generated by Score-CAM provides a more precise feature map. In our study, we aim to benefit from both gradient-guided CAM and Score-CAM and propose a gradient-guided Score-CAM in order to generate more precise saliency maps based on both fMRI and sMRI modalities. The common brain regions identified by both modalities corroborate findings from previous studies, further validating the reliability and interpretability of our framework.


This paper is organized as follows: Section \ref{sec:method} gives an overview of the designs of the proposed framework. Section \ref{sec:experiment} contains experiments on various datasets, comparing its performance and interpretability with other multi-modal fusion methods. Section \ref{sec:discussion} examines the schizophrenia-related brain functional networks and structural regions identified by our framework, discussing its limitations and outlining potential avenues for future work. Section \ref{sec:conclusion} summarizes our main contributions and findings.

\section{Methodology}\label{sec:method}

\subsection{Preliminary and Motivation}

\begin{figure*}[t]
    \centering
    \includegraphics[width=1\textwidth]{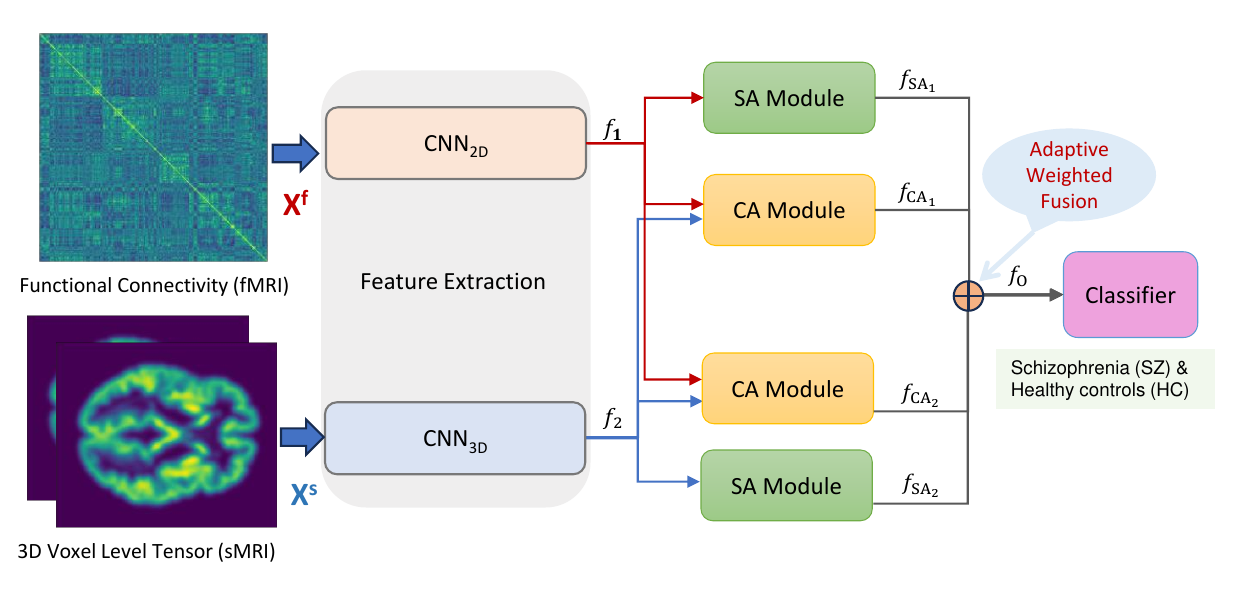}
    \vspace{-6mm}
    \caption{Overview of the proposed framework. The backbones consist of two CNN modules to extract features from the fMRI and sMRI data. Then two self-attention (SA) modules and two cross-attention (CA) modules fuse the features at the first level. The latent features are then combined by the optimal weights and input to a classifier.}
    \label{fig:fig_framework}
\end{figure*}

The objective in diagnosing schizophrenia involves classifying a population into schizophrenia (SZ) patients or healthy controls (HC), which is a typical binary classification task. Each subject sample has both fMRI and sMRI data. For fMRI, we first applied the Power atlas \cite{power2011functional} at the voxel-level, producing time-series data across 264 regions of interest (ROIs). Then we derived the functional connectivity (FC) matrix via Pearson correlation between the 264 ROIs and used the FC as the input, denoted as $\mathbf{X}^{f}_{i} \in \mathbb{R}^{264 \times 264}$ where $i$ is the index of the sample. For the sMRI modality, we directly used the 3D voxel-level data as the input, denoted as $\mathbf{X}^{s}_{i} \in \mathbb{R}^{121 \times 145 \times 121}$. Thus, the whole classification task can be defined as $\mathcal{D}_{tr} = \{\mathbf{X}^{f}_{i}, \mathbf{X}^{s}_{i}, \emph{Y}_i\}_{i=1}^N$ where $N$ is the sample size. For this binary classification task, the output $\emph{Y}_i$ is a one-hot vector, with an entry for HC and SZ subjects, respectively.

To harness the benefits of multi-modal data, we propose a novel Cross-Attentive Multi-modal Fusion (CAMF) framework to capture the interaction both within and across modalities. The proposed framework is generic and can be adapted to other types of data and multi-class classification tasks.

\subsection{Cross-Attentive Multi-modal Fusion}

\subsubsection{Framework overview.} The overall architecture of our proposed framework is shown in Figure~\ref{fig:fig_framework}. Specifically, a subject-specific 2D FC derived from fMRI is used as the input for a 2D CNN, where each entry of FC represents the connection between a pair of brain regions of interest (ROI). For sMRI, voxel-level structural MRI data are used as the input for a 3D CNN.


\subsubsection{Feature Extraction from Multi-modal MRIs}

Our proposed framework employed CNNs as feature extractors since they are ideally suited for handling the high-dimensional 2D/3D data typical of fMRI FC matrices and the volumetric data from sMRI. The FC derived from fMRI is a symmetric 2D matrix, where the order of rows and columns reflects the spatial relationship between ROIs. A global pooling layer is applied to compress the feature map of each channel and generate vectorized latent features, which are denoted as:
\begin{equation}
\mathbf{f}_1=\mathbf{GP}(\mathbf{CNN_{2D}}(\mathbf{X}^{f})),
\end{equation}
where $\mathbf{f}_1 \in \mathbb{R}^{d_1}$; $\mathbf{GP}(\cdot)$ represents the global pooling layer and $d_1=256$ is the number of channels of the last convolutional layer. For the sMRI modality, we utilize voxel-level images that contain anatomically-related information. A 3D CNN is used as the backbone and complemented by a global pooling layer, producing the following features:
\begin{equation}
\mathbf{f}_2=\mathbf{GP}(\mathbf{CNN_{3D}}(\mathbf{X}^{s})),
\end{equation}
where $\mathbf{f}_2 \in \mathbb{R}^{d_1}$. In this paper, we use global max pooling after the last convolutional layer to compress the feature maps of each channel.

CNNs maintain the spatial hierarchy of the data and offer advantages like transfer learning and the use of pre-trained models, which can benefit future researches on similar data modalities.

\subsubsection{Intra- and Inter-Modality Interaction}


To capture both intra-modal and inter-modal interactions, we utilize two fusion mechanisms, each integrating four branches of attention modules. This strategy allows us to incorporate the interactions between the two modalities while uncovering more intricate interactions among ROIs with improved interpretability.

The two self-attention (SA) modules focus on exploring interactions among subjects within individual modalities, i.e., revealing intra-modal relationships among subjects. Conversely, the cross-attention (CA) modules are tailored to explore the interactions between the two different modalities, thereby addressing the inter-modal interactions.


In each SA module, Query $\boldsymbol{q}_{i} = \boldsymbol{f}_i W_{q_i}$, Key $\boldsymbol {k}_i = \boldsymbol{f}_i W_{k_i} $ and Value $\boldsymbol {v}_i = \boldsymbol {f}_i W_{v_i}$ are generated from the same latent feature $\boldsymbol{f}_i, i \in \{1, 2\}$ where $W_{q_i}, W_{k_i}, W_{v_i} \in \mathbb{R}^{d_1 \times d_2}$ are learnable parameters and $\boldsymbol{q}_{i}, \boldsymbol{k}_i, \boldsymbol{v}_i \in \mathbb{R}^{n \times 256}$. In our experiments, we set $d_2=256$. Then the output modality-specific feature is
\begin{equation}
\boldsymbol{f}_{\text{SA}_{i}} = \mathsf{softmax}(\frac{\boldsymbol{q}_i \boldsymbol{k}_i^\top}{\sqrt{d_2}}) \boldsymbol{v}_i.
\end{equation}

For the cross-modal fusion scenario, we learn the interaction features between both modalities in the following way:
\begin{align}
\boldsymbol{f}_{\text{CA}_1} = \mathsf{softmax}(\frac{\mathbf{q}_1 \boldsymbol{k}_2^\top}{\sqrt{d_2}}) \boldsymbol{v}_1,\\
\boldsymbol{f}_{\text{CA}_2} = \mathsf{softmax}(\frac{\mathbf{q}_2 \boldsymbol{k}_1^\top}{\sqrt{d_2}}) \boldsymbol{v}_2.
\end{align}

In the second CA module, we reverse the order of modalities to address the issue of asymmetry, enabling a more effective exploration of the interaction between the two modalities.


\subsubsection{Adaptive Cross-Modal Fusion}\label{subsec:adaptive_ca_fusion}

To fuse the four outputs generated from the previous interaction module, we investigate an adaptive fusion strategy that automatically identifies the contribution of each branch. The output features from the four attention modules $\boldsymbol{f}_{\text{SA}_1}, \boldsymbol{f}_{\text{SA}_2}, \boldsymbol{f}_{\text{CA}_1}, \boldsymbol{f}_{\text{CA}_2} \in \mathbb{R}^{d}$ are subsequently fused by weighted sum
\begin{equation}
\boldsymbol{f}_O = \alpha_1 \boldsymbol{f}_{\text{SA}_1} + \alpha_2  \boldsymbol{f}_{\text{SA}_2} + \alpha_3 \boldsymbol{f}_{\text{CA}_1} + \alpha_4 \boldsymbol{f}_{\text{CA}_2},
\end{equation}
where $\alpha_{1-4}$ are learnable parameters that sum to be $1$, achieved through the $\mathsf{softmax}(\cdot)$ operation. This fusion process enables the model to learn an optimal combination of the four attention branches. The fused feature $\boldsymbol{f}_O$ is then input to the classifier.

\subsection{Objective Function and Optimization}
\label{subsec:loss_optimization}


Given the fused feature $\boldsymbol{f}_O$, a final prediction vector with two entries is calculated using a multilayer perceptron (MLP) classifier $\mathcal{C}(\cdot)$. Then we deploy the cross-entropy loss to guide model training with ground-truth supervision $\mathbf{Y}$. As a loss function, we use the standard cross-entropy loss, defined by
\begin{equation}
    \ell_{ce} = \mathsf{CrossEntropy}(\mathcal{C}(\boldsymbol{f}_O), \mathbf{Y}).
\end{equation}







We employed the Adam optimizer with a weight decay rate of 1$\mathbf{e}$-4 to optimize the model parameters. We adopted the Kaiming initialization described in \cite{he2015delving} to initialize the model parameters.


\section{Experiments}\label{sec:experiment}

\subsection{Dataset}

\subsubsection{Combined Dataset}\label{subsec:combined_data}

The evaluation is assessed using the samples combined from three datasets: Centers of Biomedical Research Excellence (COBRE) \cite{aine2017multimodal}, The Function Biomedical Informatics Research Network (FBIRN) \cite{keator2016function} and Maryland Psychiatric Research Center (MPRC) \cite{adhikari2019functional}. Together, these datasets comprise 356 healthy control (HC) samples and 264 schizophrenia (SZ) samples. Table~\ref{tab:sample_size} lists the detailed sample size of each subset.

\begin{table}[htb!]
\linespread{2}
\renewcommand\arraystretch{1.3}
  \begin{center}
    \caption{Statistic of the datasets.}\vspace{-5mm}
    \label{tab:sample_size}
    \begin{tabular}{ccc}\\
        \toprule
        Dataset & SZ samples & HC samples \\
        \midrule
        COBRE & 57 & 81 \\
        FBIRN & 127 & 152\\
        MPRC & 80 & 123\\\hline
        BSNIP & 199 & 243\\
        \midrule
        Total & 463 & 599\\
        \bottomrule
    \end{tabular}
  \end{center}
\end{table}

\subsubsection{Bipolar and Schizophrenia Network for Intermediate Phenotypes (BSNIP)}\label{subsec:BSNIP_data}

We conducted independent testing on datasets with different sources and sampling protocols for the generalizability of the framework. Herein, we used the Bipolar and Schizophrenia Network for Intermediate Phenotypes (BSNIP) dataset to validate the performance.


\subsubsection{Data preprocessing}\label{subsec:data_preprocessing}

We follow the same standard pipeline for MRI data pre-processing as described in \cite{rahaman2023deep}. Specifically, the sMRI scans are preprocessed using the statistical parametric mapping (SPM12) toolbox\footnote{ \url{https://www.fil.ion.ucl.ac.uk/spm/software/spm12/}}. The unified segmentation and normalization were applied to the sMRI scans for gray matter, white matter, and cerebrospinal fluid (CSF), and a modulated normalization algorithm was used to generate gray matter volume (GMV) and gray matter density (GMD). Then the GMV and GMD were smoothed using a Gaussian kernel with a full width at half maximum (FWHM) = 6 mm.

The fMRI data were preprocessed using the SPM toolbox within MATLAB 2020b. The first five scans were removed for the signal equilibrium and participants' adaptation to the scanner's noise. We performed rigid body motion correction using the toolbox in SPM to correct subject head motion, followed by the slice-timing correction to account for timing differences in slice acquisition. The fMRI data were subsequently normalized into the standard Montreal Neurological Institute (MNI) space using an echo-planar imaging (EPI) template and were slightly resampled to $3 \times 3 \times 3$ mm$^3$ isotropic voxels. The resampled fMRI images were further smoothed using a Gaussian kernel with a FWHM = 6 mm.

\subsection{Comparison Experiment}\label{subsec:exp_set-up}

\subsubsection{Cross Validation} In this study, we adopt 5-fold cross-validation to evaluate the model performance. The patient index codes are shuffled and divided evenly into 5 folds, and one fold was left out as the test set. For the rest 4 folds, we select 1/8 to be the validation set and the larger portion to be the training set. The averaged results of 5 folds are presented for all experiments.



\subsubsection{Evaluation Metrics}

An issue with the combined dataset is the imbalance in disease class. To address this concern, we employ more reliable evaluation metrics, such as F1 score and Matthew's correlation coefficient (MCC), alongside the standard measure of classification accuracy.

The F1 score is calculated as:
\begin{equation}
\text{F}_1 = \frac{2 * \rm{TP}}{2 * \rm{TP} + \rm{FP} + \rm{FN}},
\end{equation}
where the minimum value is $0$ and the maximum value is $+1$. 

The MCC is defined as 
\begin{equation}\small
\rm{MCC} = \frac{\rm{TP} * \rm{TN} - \rm{FP} * \rm{FN}}{\sqrt{(\rm{TP}+\rm{FP}) * (\rm{TP}+\rm{FN}) *(\rm{TN}+\rm{FP}) *(\rm{TN}+\rm{FN})}},
\end{equation}
where the minimum value is $-1$ and the maximum value is $+1$. MCC is recommended by the National Institutes of Health (NIH) \cite{chicco2020advantages} to be a reliable statistic because it produces a high score when the prediction obtained good results in all of the four confusion matrix categories (true positives: TP, false negatives: FN, true negatives: TN, and false positives: FN).


\begin{table*}[t]
\linespread{2} 
\renewcommand\arraystretch{1.3}
  \caption{The prediction performance for all multi-modal fusion methods under comparison}
  \label{tab:results}
  \centering
  \begin{tabular}{cccccccc}
    \toprule
    Dataset & Exp ID & Feature Extractor & Fusion Method & Classifier & F1-Score & Acc & MCC-Score \\
    \midrule
     \multirow{5}{*}{\makecell{COBRE\\+FBIRN\\+MPRC}}
    & 1 & PCA & Element-wise Sum & SVM & $0.6355\pm0.0377$ & $0.7048\pm0.0333$ & $0.3968\pm0.0630$\\
    & 2 & PCA & Element-wise Sum & SGDClassifier & $0.6336\pm0.0453$ & $0.7065\pm0.0309$ & $0.3979\pm0.0611$\\
    & 3 & PCA & Element-wise Sum & MLP & $0.4529\pm0.2279$ & $0.4952\pm0.0613$ & $0.0450\pm0.1228$\\
    & 4 & CNNs & Element-wise Sum & MLP & $0.6469\pm0.1031$ & $0.7274\pm0.0681$ & $0.4284\pm0.1441$\\
    & 5 & CNNs & Concatenation & MLP & $0.6173\pm0.1457$ & $0.7113\pm0.0496$ & $0.4296\pm0.0799$\\
    & 6 & CNNs & \textbf{CAMF} &  MLP  & $0.7049\pm0.0159$ & $0.7435\pm0.0129$ & $0.4906\pm0.0303$\\\hline
     \multirow{3}{*}{BSNIP}
    & 7 & CNNs & Element-wise Sum & MLP & $0.6460\pm0.0454$ & $0.6629\pm0.0318$ & $0.3307\pm0.0625$\\
    & 8 & CNNs & Concatenation &  MLP  & $0.6285\pm0.0569$ & $0.6697\pm0.0262$ & $0.3536\pm0.0458$\\
    & 9 & CNNs & \textbf{CAMF} &  MLP  & $0.6843\pm0.0259$ & $0.6787\pm0.0183$ & $0.3614\pm0.0373$\\
    \bottomrule
  \end{tabular}
\end{table*}

\subsubsection{Baselines} We compare the combination of different feature extractors, fusion methods and classifiers, which demonstrates the superiority of our proposed model.

For feature extractors, we compare principal component analysis (PCA) and CNNs. For PCA, we vectorize the inputs for both modalities, and apply PCA to reduce the dimension to $d=256$, matching our framework's latent feature size. To ensure consistency of principal components between the training and test sets, we first perform PCA on the training set and then apply this transformation matrix to the test set.


Regarding fusion methods, we explore element-wise sum operation, concatenation, and our proposed CAMF; for classifiers, we adopt Support Vector Machine (SVM) \cite{cortes1995support}, linear SVM with stochastic gradient descent learning (SGDClassifier) \cite{amari1993backpropagation}, and MLP, which are commonly used linear classifiers.

For a fair comparison of classification performance between our proposed framework and other multi-modal fusion methods, we ensure that the structure of CNN backbones and the MLP classifier align with our proposed architecture. Such comparison avoids the bias caused by differences in backbones and the final classifier.

\subsubsection{Results and Discussion} The detailed experimental results are shown in Table~\ref{tab:results}. Specifically, we have the following conclusions:
\begin{itemize}
    \item \textbf{Comparing CAMF with linear classifiers:}
    Comparing Exp 1, 2, and 6, we can see that our proposed framework significantly outperforms all three baseline linear methods according to all evaluation metrics, especially for MCC. This shows the advantage of deep neural networks over linear models.
    \item \textbf{Comparing MLP with linear classifiers:} The comparison between Exp 1-3 shows that MLP does not work well with PCA as the backbone to extract latent features. A reason is that feature selection via PCA and classification with MLP are conducted as distinct, separate processes. 
    \item \textbf{Comparing CNNs with PCA:}
    The comparison between Exp 3 and 4 shows that PCA's effectiveness is compromised by the imbalanced distribution of samples across two classes and PCA cannot capture the essential patterns in the data compared to CNNs as backbones.
    \item \textbf{Comparing CAMF with simple data fusions:} The results of Exp 4-6 indicate that the fusion of latent features from multiple modalities may lower the performance if not paired with an appropriate data fusion. Simple concatenation and element-wise sum cannot combine the information from both modalities. The superior performance of CAMF, compared to other fusion methods, demonstrates its ability to extract and merge both inter-modal and intra-modal interactions, resulting in significant improvement. The independent experiments on the BSNIP dataset (Exp 7-9) further corroborate these findings, thereby reinforcing the validity of our conclusions.
\end{itemize}

\subsection{Ablation Study}
To evaluate the impact of each component in our proposed framework, we conducted the ablation study. Specifically, we assess the importance of each component by observing the changes in performance resulting from its removal. Here we use the CNN as feature extractors for all scenarios and examine the impact of the two-level data fusion (attention modules and adaptive weights). The results of all scenarios are shown in Table \ref{tab:ablation_results}, leading to the following conclusions:
\begin{itemize}
    \item \textbf{Uni-modal input of fMRI (Exp 1, 10) or sMRI (Exp 2, 11):} The proposed CAMF framework significantly exceeds the uni-modal methods with respect to all evaluation metrics. This indicates that CAMF can effectively combine the complementary information of multiple modalities to improve classification performance.
    \item \textbf{Simple fusion methods without attention modules (Exp 3-5, 12-14):} Similar to the baselines, the better performance of CAMF displays the advantage of the two-level cross-attentive fusion method proposed in our CAMF framework. Compared to only using the adaptive weights (Exp 5, 14) for the fusion, the results of CAMF underscore the enhanced predictive capability brought by the SA and CA modules.
    \item \textbf{Removing cross-attention or self-attention modules (Exp 5-7, 14-16):} The experimental results show that the cross-attention module (Exp 7, 16) can significantly improve the performance than without using it (Exp 5, 14). This indicates that the CA modules could well exploit the interaction between modalities, thereby leading to better classification results. Although employing only the SA modules (Exp 6, 15) can not independently improve the performance, their combination with the CA modules proves to be beneficial. This proves that both CA and SA are indispensable components of the proposed method.
    \item \textbf{Removing adaptive weights (Exp 8, 17):} Finally, we compare the fusion method with adaptive weights with simple concatenation (Exp 8, 17). This shows that the fusion with optimal weight can effectively incorporate the information from inter- and within-modality interactions. 
          
\end{itemize}


The ablation study shows the importance of each modality and underscores the contributions of each component within the proposed framework. However, it cannot give a human-comprehensible level interpretation of the model behaviors, such as at the region level or even pixel/voxel level. Thus, we further utilized the Score-CAM method to generate high-resolution saliency maps that help to understand the model's focus and identify biomarkers associated with schizophrenia.

\begin{table*}[t]
\linespread{2} 
\renewcommand\arraystretch{1.3}
  \caption{The prediction results of the ablation study.}
  \label{tab:ablation_results}
  \centering
  \begin{tabular}{ccccccc}
    \toprule
    Dataset & Exp ID& Modality Type & Fusion Method & F1-Score & Acc & MCC-Score \\
    \midrule
     \multirow{13}{*}{\makecell{COBRE\\+FBIRN\\+MPRC}}
    & 1 & Uni-modal (fMRI) & - & $0.5181\pm0.0479$ & $0.6355\pm0.0219$ & $0.2519\pm0.0530$\\
    & 2 & Uni-modal (sMRI) & - & $0.6384\pm0.0491$ & $0.7129\pm0.0341$ & $0.4110\pm0.0589$\\
    & 3 & Multi-modal & Element-wise Sum & $0.6469\pm0.1031$ & $0.7274\pm0.0681$ & $0.4284\pm0.1441$\\
    & 4 & Multi-modal & Concatenation & $0.6173\pm0.1457$ & $0.7113\pm0.0496$ & $0.4296\pm0.0799$\\
    & 5 & Multi-modal & Adaptive Weights & $0.6278\pm0.0732$ & $0.7032\pm0.0644$ & $0.4123\pm0.1258$\\
    & 6 & Multi-modal & \makecell{2 Self-attention\\+ Adaptive Weights} & $0.6128\pm0.0403$ & $0.6645\pm0.0464$ & $0.3489\pm0.0626$\\
    & 7 & Multi-modal & \makecell{2 Cross-attention\\+ Adaptive Weights} & $0.6846\pm0.0297$ & $0.7435\pm0.0344$ & $0.4767\pm0.0557$\\
    & 8 & Multi-modal & \makecell{2 Cross-attention\\+ 2 Self-attention\\+ Concatenation} & $0.6379\pm0.0339$ & $0.7097\pm0.0177$ & $0.4078\pm0.0334$\\
    & 9 & \textbf{Multi-modal} & \textbf{CAMF} & \boldmath$0.7049\pm0.0159$\unboldmath & \boldmath $0.7435\pm0.0129$ \unboldmath & \boldmath$0.4906\pm0.0303$\unboldmath\\\hline
     \multirow{13}{*}{BSNIP}
    & 10 & Uni-modal (fMRI) & - & $0.5378\pm0.1031$ & $0.6045\pm0.0802$ & $0.2225\pm0.1633$\\
    & 11 & Uni-modal (sMRI) & - & $0.6425\pm0.0425$ & $0.6337\pm0.0208$ & $0.2748\pm0.0498$\\
    & 12 & Multi-modal & Element-wise Sum & $0.6460\pm0.0454$ & $0.6629\pm0.0318$ & $0.3307\pm0.0625$\\
    & 13 & Multi-modal & Concatenation & $0.6285\pm0.0569$ & $0.6697\pm0.0262$ & $0.3536\pm0.0458$\\
    & 14 & Multi-modal & Adaptive Weights & $0.6328\pm0.0461$ & $0.6517\pm0.0100$ & $0.3112\pm0.0174$\\
    & 15 & Multi-modal & \makecell{2 Self-attention\\+ Adaptive Weights} & $0.6379\pm0.1198$ & $0.6449\pm0.0453$ & $0.3052\pm0.0814$\\
    & 16 & Multi-modal & \makecell{2 Cross-attention\\+ Adaptive Weights} & $0.6767\pm0.0414$ & $0.6809\pm0.0323$ & $0.3662\pm0.0670$\\
    & 17 & Multi-modal & \makecell{2 Cross-attention\\+ 2 Self-attention\\+ Concatenation} & $0.6655\pm0.0363$ & $0.6742\pm0.0275$ & $0.3510\pm0.0552$\\
    & 18 & \textbf{Multi-modal} & \textbf{CAMF} & \boldmath$0.6843\pm0.0259$\unboldmath & \boldmath$0.6787\pm0.0183$\unboldmath & \boldmath$0.3614\pm0.0373$\unboldmath\\
    \bottomrule
  \end{tabular}
\end{table*}


\subsection{Score-CAM Based Interpretation}

Score-CAM \cite{wang2020score} is an improvement over the class activation map (CAM) \cite{zhou2016learning} method, which generates the saliency map and highlights important features or brain regions when making the prediction with the model. In our case of Schizophrenia diagnosis, the highlighted regions can be interpreted as the key disease-related brain functional networks (BFN) and brain structural regions (BSR).

\begin{figure}[htp!]
    \centering
    \includegraphics[width=0.495\textwidth]{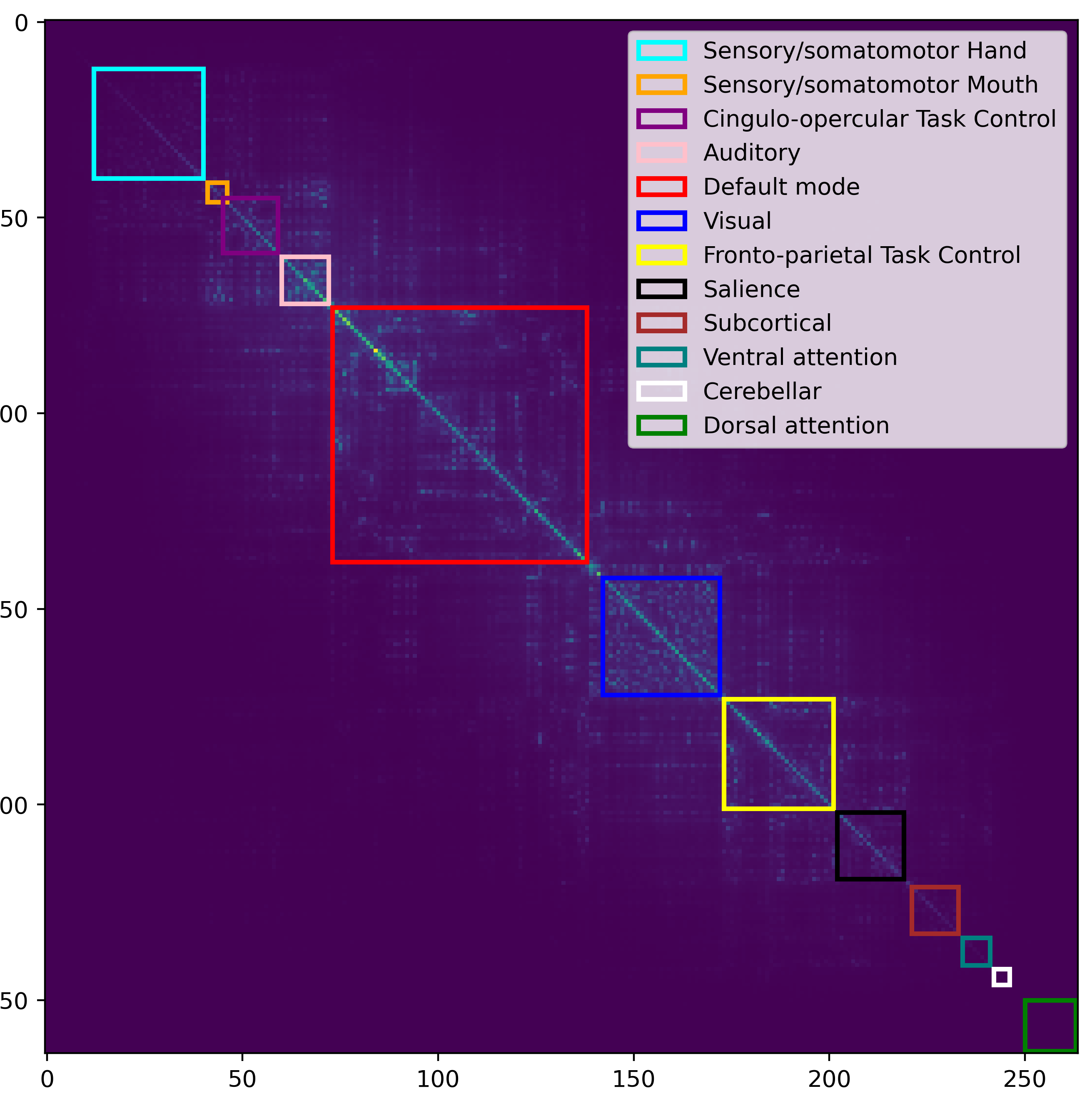}
    \caption{Average saliency map from the FC modality, where the rows and columns in the FC matrix are grouped by different functional brain regions, which are highlighted in various colored boxes. We observe a high correlation with each box.}
    \label{fig:fc_saliency_map}\vspace{-0mm}
\end{figure}

Using the Score-CAM method at the last convolutional layer of each backbone, we produced two saliency maps for each subject with  low resolutions ($6 \times 6$ for fMRI and $5 \times 7 \times 5$ for sMRI) due to the pooling layers in the backbones. Then we interpolate the saliency maps to match with the size of the input data ($264 \times 264$ for fMRI and $121 \times 145 \times 121$ for sMRI). The high-resolution saliency maps after interpolation allow biomarker identification from the input data. In order to identify the disease-related BFNs and BSRs among the whole population, for each modality, we generated an average template by averaging the saliency maps of all subjects. Like the gradient-guided CAM method \cite{selvaraju2017grad}, we also apply element-wise multiplication of gradients of the predicted class to the saliency maps to generate the refined regions.

For the fMRI, the input FC is always symmetric since it's derived by Pearson correlation. However, the saliency map could be asymmetric due to the padding of kernels in CNNs. Thus we show the average of FC saliency map template and its transpose to address the asymmetry issue. The results are shown in Fig \ref{fig:fc_saliency_map}.

\begin{figure*}[htp]
    \centering
    \includegraphics[width=0.995\textwidth]{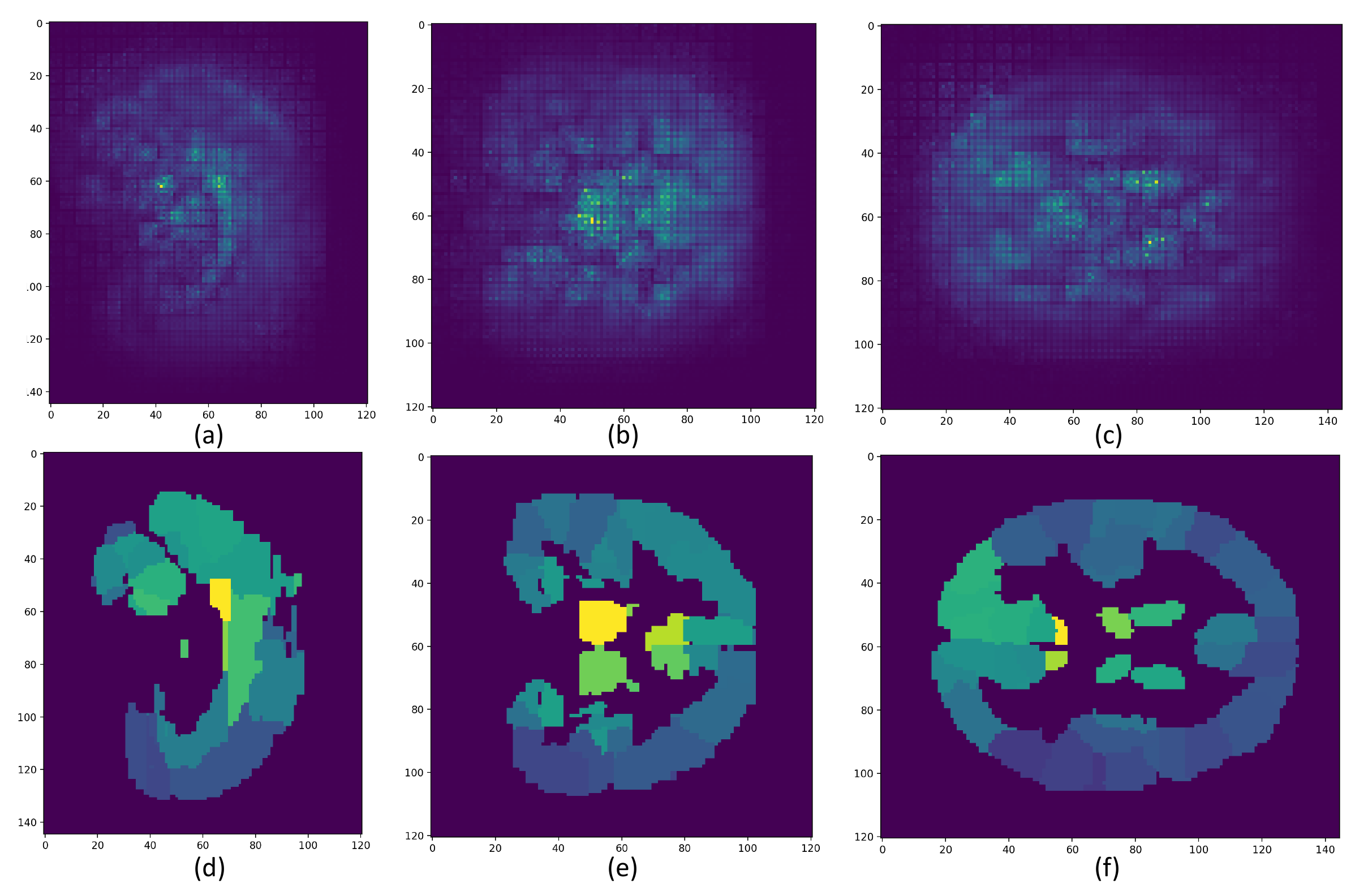}\vspace{-3mm}
    \caption{Voxel-level and region-level Score-CAM saliency maps from \emph{x}-, \emph{y}- and \emph{z}- axis. In the region-level saliency maps the intensity is averaged inside each brain region defined by the automated anatomical labeling (AAL) \cite{tzourio2002automated} atlas. Subfigure (a)-(c) are the voxel-level average saliency map and subfiture (d)-(f) are the region-level average saliency map. For the subfigure of each axis, we choose the slice in the middle of both poles to visualize the highlighted voxels/regions in the center of the brain.}\vspace{-3mm}
  \label{fig:smri_saliency_map}
\end{figure*}

For the sMRI, we first visualize the voxel-level average saliency map template from the x-, y- and z-axis in Fig \ref{fig:smri_saliency_map} (a)-(c). In order to locate the disease-related BSRs, we then segment the template using the automated anatomical labeling (AAL) \cite{tzourio2002automated} atlas. The voxels in each anatomical BSR are grouped together and the BSRs are ranked by their averaged activation scores. The region-level saliency map template is shown in Fig \ref{fig:smri_saliency_map} (d)-(f).

\section{Discussion}\label{sec:discussion}

\subsection{Schizophrenia-related Brain Region Identification}


The saliency map template from the fMRI shows the key disease-related pair-wise functional connectivities. According to Fig \ref{fig:fc_saliency_map}, most highlighted BFNs lie in the auditory, default mode, and visual networks. The high activation scores link the connections within these regions to Schizophrenia, which aligns with the results from previous research. The auditory network is considered responsible for receiving and processing sound, which is the basis of comprehension and analysis of the meaning of utterances \cite{hackett2015anatomic}. E.g., Joo et al. (2020) \cite{joo2020aberrant} revealed a significantly decreased FC in the right postcentral gyrus (the auditory network) of Schizophrenia patients compared to healthy controls. The visual network is involved with processing visual information and the reduced intrinsic visual cortical connectivity within it was found to be associated with impaired perceptual closure in schizophrenia by a previous study \cite{van2017reduced}. The default mode network is a set of brain regions that become more active when the individual is in a resting state \cite{buckner2012serendipitous}. A study \cite{garrity2007aberrant} concluded that in the default mode network, both temporal frequency alterations and disruption of local spatial patterns are associated with schizophrenia.

The saliency map template of sMRI in Fig \ref{fig:smri_saliency_map}(a)-(c) shows the voxel-level intensity in the 3D space that the model utilizes for the classification. By ranking the BSRs according to their average activation scores, the top three highlighted BSRs are cingulum, thalamus, and caudate. The cingulum bundle is one of the most distinctive fiber tracts in the brain and several prior studies \cite{fitzsimmons2020cingulum, whitford2014localized} revealed the link between cingulum bundle abnormalities and schizophrenia. The thalamus is primarily a gray matter structure within the diencephalon, which plays a key role in linking and relaying information between brain regions, including the visual system and the primary auditory cortex. This matches the disease-related BFNs identified by the FC. Prior research also found the thalamus to be a region associated with schizophrenia \cite{pergola2015role}. The caudate nucleus plays a critical role in several higher cognitive functions, such as memory, language, and emotion. A prior study \cite{takase2004reduced} identified the caudate to be associated with schizophrenia since the caudate nucleus of the patients have smaller volumes compared to the healthy controls.

The spatial overlaps between BFNs (auditory, default mode, and visual networks) identified from fMRI and BSRs (cingulum, thalamus, and caudate) identified from sMRI validated the reliability and interpretability of our proposed framework.

\subsection{Limitations and Future Work}

Although our proposed framework successfully improved the classification performance by fusing multi-modal data effectively, there are still limitations to address and potential avenues for future research. First, our framework deployed a 2D CNN as the backbone to extract latent features based on the FCs. Employing a graph-based representation for FCs could better capture structural information, potentially elevating both performance and interpretability. Second, although Score-CAM can produce saliency maps on CNN backbones, other interpretable methods can be exploited for potential improvements. Third, while our proposed framework is tested here for the two modalities, it can be generalized to three or more modalities (e.g., SNPs), which can further improve the predictive performance.

\subsection{Data Availability}

All the datasets are available online and can be requested from the corresponding authors in the referred studies. The data preprocessing pipeline is publicly available at \url{http://trendscenter.org/software} and the code of models is available online at \url{https://github.com/zzyBen/TMI_cross_attentive_multimodal_MRI/tree/main}.

\section{Conclusion}\label{sec:conclusion}
Our proposed framework notably improved classification ability and provided interpretable results via fusing fMRI and sMRI data. Drawing inspiration from the transformer model's success, we introduced self-attention and cross-attention modules to incorporate the interactions within and between each modality, leading to a better model to represent multi-modal data. The experimental results on multiple datasets demonstrated its superiority over conventional learning methods and multi-modal fusion methods. In addition, Score-CAM was used to identify critical disease-related functional networks and structural regions. These findings are in concordance with the results reported in prior studies, validating the framework in biomarker discovery and disease diagnosis.


\balance
\bibliography{Reference.bib}

\begin{thebibliography}{10}
\providecommand{\url}[1]{#1}
\csname url@samestyle\endcsname
\providecommand{\newblock}{\relax}
\providecommand{\bibinfo}[2]{#2}
\providecommand{\BIBentrySTDinterwordspacing}{\spaceskip=0pt\relax}
\providecommand{\BIBentryALTinterwordstretchfactor}{4}
\providecommand{\BIBentryALTinterwordspacing}{\spaceskip=\fontdimen2\font plus
\BIBentryALTinterwordstretchfactor\fontdimen3\font minus \fontdimen4\font\relax}
\providecommand{\BIBforeignlanguage}[2]{{%
\expandafter\ifx\csname l@#1\endcsname\relax
\typeout{** WARNING: IEEEtran.bst: No hyphenation pattern has been}%
\typeout{** loaded for the language `#1'. Using the pattern for}%
\typeout{** the default language instead.}%
\else
\language=\csname l@#1\endcsname
\fi
#2}}
\providecommand{\BIBdecl}{\relax}
\BIBdecl

\bibitem{oh2020identifying}
J.~Oh, B.-L. Oh, K.-U. Lee, J.-H. Chae, and K.~Yun, ``Identifying schizophrenia using structural mri with a deep learning algorithm,'' \emph{Frontiers in psychiatry}, vol.~11, p.~16, 2020.

\bibitem{katuwal2015predictive}
G.~J. Katuwal, N.~D. Cahill, S.~A. Baum, and A.~M. Michael, ``The predictive power of structural mri in autism diagnosis,'' in \emph{2015 37th annual international conference of the ieee engineering in medicine and biology society (EMBC)}, 2015, pp. 4270--4273.

\bibitem{tomassini2021end}
S.~Tomassini, N.~Falcionelli, P.~Sernani, H.~M{\"u}ller, and A.~F. Dragoni, ``An end-to-end 3d convlstm-based framework for early diagnosis of alzheimer's disease from full-resolution whole-brain smri scans,'' in \emph{2021 IEEE 34th International Symposium on Computer-Based Medical Systems (CBMS)}, 2021, pp. 74--78.

\bibitem{belliveau1991functional}
J.~W. Belliveau \emph{et~al.}, ``Functional mapping of the human visual cortex by magnetic resonance imaging,'' \emph{Science}, vol. 254, no. 5032, pp. 716--719, 1991.

\bibitem{li2018brain}
H.~Li, T.~D. Satterthwaite, and Y.~Fan, ``Brain age prediction based on resting-state functional connectivity patterns using convolutional neural networks,'' in \emph{2018 ieee 15th international symposium on biomedical imaging (isbi 2018)}, 2018, pp. 101--104.

\bibitem{zhang2017hybrid}
Y.~Zhang, H.~Zhang, X.~Chen, S.-W. Lee, and D.~Shen, ``Hybrid high-order functional connectivity networks using resting-state functional mri for mild cognitive impairment diagnosis,'' \emph{Scientific reports}, vol.~7, no.~1, p. 6530, 2017.

\bibitem{wang2021functional}
J.~Wang \emph{et~al.}, ``Functional network estimation using multigraph learning with application to brain maturation study,'' \emph{Human brain mapping}, vol.~42, no.~9, pp. 2880--2892, 2021.

\bibitem{finn2015functional}
E.~S. Finn, X.~Shen, D.~Scheinost, M.~D. Rosenberg, J.~Huang, M.~M. Chun, X.~Papademetris, and R.~T. Constable, ``Functional connectome fingerprinting: identifying individuals using patterns of brain connectivity,'' \emph{Nature neuroscience}, vol.~18, no.~11, pp. 1664--1671, 2015.

\bibitem{sui2013three}
J.~Sui \emph{et~al.}, ``Three-way (n-way) fusion of brain imaging data based on mcca+ jica and its application to discriminating schizophrenia,'' \emph{NeuroImage}, vol.~66, pp. 119--132, 2013.

\bibitem{zu2016label}
C.~Zu, B.~Jie, M.~Liu, S.~Chen, D.~Shen, D.~Zhang, and A.~D.~N. Initiative, ``Label-aligned multi-task feature learning for multimodal classification of alzheimer’s disease and mild cognitive impairment,'' \emph{Brain imaging and behavior}, vol.~10, pp. 1148--1159, 2016.

\bibitem{yang2023mapping}
Y.~Yang, C.~Ye, X.~Guo, T.~Wu, Y.~Xiang, and T.~Ma, ``Mapping multi-modal brain connectome for brain disorder diagnosis via cross-modal mutual learning,'' \emph{IEEE Transactions on Medical Imaging}, 2023.

\bibitem{hu2019deep}
W.~Hu, B.~Cai, A.~Zhang, V.~D. Calhoun, and Y.-P. Wang, ``Deep collaborative learning with application to the study of multimodal brain development,'' \emph{IEEE Transactions on Biomedical Engineering}, vol.~66, no.~12, pp. 3346--3359, 2019.

\bibitem{hu2021interpretable}
W.~Hu \emph{et~al.}, ``Interpretable multimodal fusion networks reveal mechanisms of brain cognition,'' \emph{IEEE transactions on medical imaging}, vol.~40, no.~5, pp. 1474--1483, 2021.

\bibitem{vaswani2017attention}
A.~Vaswani, N.~Shazeer, N.~Parmar, J.~Uszkoreit, L.~Jones, A.~N. Gomez, {\L}.~Kaiser, and I.~Polosukhin, ``Attention is all you need,'' \emph{Advances in neural information processing systems}, vol.~30, 2017.

\bibitem{qu2023interpretable}
G.~Qu, A.~Orlichenko, J.~Wang, G.~Zhang, L.~Xiao, K.~Zhang, T.~W. Wilson, J.~M. Stephen, V.~D. Calhoun, and Y.-P. Wang, ``Interpretable cognitive ability prediction: A comprehensive gated graph transformer framework for analyzing functional brain networks,'' \emph{IEEE Transactions on Medical Imaging}, 2023.

\bibitem{zhu2022multimodal}
Q.~Zhu, H.~Wang, B.~Xu, Z.~Zhang, W.~Shao, and D.~Zhang, ``Multimodal triplet attention network for brain disease diagnosis,'' \emph{IEEE Transactions on Medical Imaging}, vol.~41, no.~12, pp. 3884--3894, 2022.

\bibitem{salahuddin2022transparency}
Z.~Salahuddin, H.~C. Woodruff, A.~Chatterjee, and P.~Lambin, ``Transparency of deep neural networks for medical image analysis: A review of interpretability methods,'' \emph{Computers in biology and medicine}, vol. 140, p. 105111, 2022.

\bibitem{huff2021interpretation}
D.~T. Huff, A.~J. Weisman, and R.~Jeraj, ``Interpretation and visualization techniques for deep learning models in medical imaging,'' \emph{Physics in Medicine \& Biology}, vol.~66, no.~4, p. 04TR01, 2021.

\bibitem{selvaraju2017grad}
R.~R. Selvaraju, M.~Cogswell, A.~Das, R.~Vedantam, D.~Parikh, and D.~Batra, ``Grad-cam: Visual explanations from deep networks via gradient-based localization,'' in \emph{Proceedings of the IEEE international conference on computer vision}, 2017, pp. 618--626.

\bibitem{wang2020score}
H.~Wang, Z.~Wang, M.~Du, F.~Yang, Z.~Zhang, S.~Ding, P.~Mardziel, and X.~Hu, ``Score-cam: Score-weighted visual explanations for convolutional neural networks,'' in \emph{Proceedings of the IEEE/CVF conference on computer vision and pattern recognition workshops}, 2020, pp. 24--25.

\bibitem{power2011functional}
J.~D. Power \emph{et~al.}, ``Functional network organization of the human brain,'' \emph{Neuron}, vol.~72, no.~4, pp. 665--678, 2011.

\bibitem{he2015delving}
K.~He, X.~Zhang, S.~Ren, and J.~Sun, ``Delving deep into rectifiers: Surpassing human-level performance on imagenet classification,'' in \emph{Proceedings of the IEEE international conference on computer vision}, 2015, pp. 1026--1034.

\bibitem{aine2017multimodal}
C.~Aine \emph{et~al.}, ``Multimodal neuroimaging in schizophrenia: description and dissemination,'' \emph{Neuroinformatics}, vol.~15, pp. 343--364, 2017.

\bibitem{keator2016function}
D.~B. Keator \emph{et~al.}, ``The function biomedical informatics research network data repository,'' \emph{Neuroimage}, vol. 124, pp. 1074--1079, 2016.

\bibitem{adhikari2019functional}
B.~M. Adhikari \emph{et~al.}, ``Functional network connectivity impairments and core cognitive deficits in schizophrenia,'' \emph{Human brain mapping}, vol.~40, no.~16, pp. 4593--4605, 2019.

\bibitem{rahaman2023deep}
M.~A. Rahaman, J.~Chen, Z.~Fu, N.~Lewis, A.~Iraji, T.~G. van Erp, and V.~D. Calhoun, ``Deep multimodal predictome for studying mental disorders,'' \emph{Human Brain Mapping}, vol.~44, no.~2, pp. 509--522, 2023.

\bibitem{chicco2020advantages}
D.~Chicco and G.~Jurman, ``The advantages of the matthews correlation coefficient (mcc) over f1 score and accuracy in binary classification evaluation,'' \emph{BMC genomics}, vol.~21, no.~1, pp. 1--13, 2020.

\bibitem{cortes1995support}
C.~Cortes and V.~Vapnik, ``Support-vector networks,'' \emph{Machine learning}, vol.~20, pp. 273--297, 1995.

\bibitem{amari1993backpropagation}
S.-i. Amari, ``Backpropagation and stochastic gradient descent method,'' \emph{Neurocomputing}, vol.~5, no. 4-5, pp. 185--196, 1993.

\bibitem{zhou2016learning}
B.~Zhou, A.~Khosla, A.~Lapedriza, A.~Oliva, and A.~Torralba, ``Learning deep features for discriminative localization,'' in \emph{Proceedings of the IEEE conference on computer vision and pattern recognition}, 2016, pp. 2921--2929.

\bibitem{tzourio2002automated}
N.~Tzourio-Mazoyer, B.~Landeau, D.~Papathanassiou, F.~Crivello, O.~Etard, N.~Delcroix, B.~Mazoyer, and M.~Joliot, ``Automated anatomical labeling of activations in spm using a macroscopic anatomical parcellation of the mni mri single-subject brain,'' \emph{Neuroimage}, vol.~15, no.~1, pp. 273--289, 2002.

\bibitem{hackett2015anatomic}
T.~A. Hackett, ``Anatomic organization of the auditory cortex,'' \emph{Handbook of clinical neurology}, vol. 129, pp. 27--53, 2015.

\bibitem{joo2020aberrant}
S.~W. Joo, W.~Yoon, Y.~T. Jo, H.~Kim, Y.~Kim, and J.~Lee, ``Aberrant executive control and auditory networks in recent-onset schizophrenia,'' \emph{Neuropsychiatric Disease and Treatment}, pp. 1561--1570, 2020.

\bibitem{van2017reduced}
V.~van~de Ven, A.~R. Jagiela, V.~Oertel-Kn{\"o}chel, and D.~E. Linden, ``Reduced intrinsic visual cortical connectivity is associated with impaired perceptual closure in schizophrenia,'' \emph{NeuroImage: Clinical}, vol.~15, pp. 45--52, 2017.

\bibitem{buckner2012serendipitous}
R.~L. Buckner, ``The serendipitous discovery of the brain's default network,'' \emph{Neuroimage}, vol.~62, no.~2, pp. 1137--1145, 2012.

\bibitem{garrity2007aberrant}
A.~G. Garrity, G.~D. Pearlson, K.~McKiernan, D.~Lloyd, K.~A. Kiehl, and V.~D. Calhoun, ``Aberrant “default mode” functional connectivity in schizophrenia,'' \emph{American journal of psychiatry}, vol. 164, no.~3, pp. 450--457, 2007.

\bibitem{fitzsimmons2020cingulum}
J.~Fitzsimmons \emph{et~al.}, ``Cingulum bundle abnormalities and risk for schizophrenia,'' \emph{Schizophrenia research}, vol. 215, pp. 385--391, 2020.

\bibitem{whitford2014localized}
T.~J. Whitford \emph{et~al.}, ``Localized abnormalities in the cingulum bundle in patients with schizophrenia: a diffusion tensor tractography study,'' \emph{NeuroImage: Clinical}, vol.~5, pp. 93--99, 2014.

\bibitem{pergola2015role}
G.~Pergola, P.~Selvaggi, S.~Trizio, A.~Bertolino, and G.~Blasi, ``The role of the thalamus in schizophrenia from a neuroimaging perspective,'' \emph{Neuroscience \& Biobehavioral Reviews}, vol.~54, pp. 57--75, 2015.

\bibitem{takase2004reduced}
K.~Takase, C.~Tamagaki, G.~Okugawa, K.~Nobuhara, T.~Minami, T.~Sugimoto, S.~Sawada, and T.~Kinoshita, ``Reduced white matter volume of the caudate nucleus in patients with schizophrenia,'' \emph{Neuropsychobiology}, vol.~50, no.~4, pp. 296--300, 2004.

\end{thebibliography}

\end{document}